\documentclass[conference]{IEEEtran}
\IEEEoverridecommandlockouts
\usepackage{cite}
\usepackage{amsmath,amssymb,amsfonts}
\usepackage{algorithmic}
\usepackage{graphicx}
\usepackage{textcomp}
\usepackage{xcolor}
\usepackage{hyperref}
\usepackage{url}
\def\BibTeX{{\rm B\kern-.05em{\sc i\kern-.025em b}\kern-.08em
    T\kern-.1667em\lower.7ex\hbox{E}\kern-.125emX}}
    
\begin{document}

\title{Online Gradient Descent for Grid Regulated Power Point Tracking Under a Highly Fluctuating Weather and
Load}

\author{
\IEEEauthorblockN{Muhy Eddin Za'ter}
\IEEEauthorblockA{\textit{Electrical Engineering Department} \\
\textit{University of Colorado Boulder}\\
Boulder, Colorado \\
muhy.zater@colorado.edu}
\and
\IEEEauthorblockN{Sandy Yacoub Miguel}
\IEEEauthorblockA{\textit{IEEE Student} \\
\textit{Princess Sumaya University for Technology}\\
Amman, Jordan \\
san20190272@std.psut.edu.jo}
\and
\IEEEauthorblockN{Majd Ghazi Batarseh}
\IEEEauthorblockA{\textit{Electrical Engineering Department} \\
\textit{Princess Sumaya University for Technology}\\
Amman, Jordan \\
m.batarseh@psut.edu.jo}
}

\maketitle

\begin{abstract}
The heavy reliance on Photovoltaic (PV) systems to meet the increasing electricity demand cost-effectively imposes new challenges on researchers and developers to guarantee the operation at maximum power point (MPP) under all conditions. However, the increasing penetration of renewable energy into the grid creates additional difficulties for the grid operators, which requires the limitation and flexibility of the energy contribution from those renewable resources to a specific grid-regulated threshold for stability and reliability purposes. 
All of these technical and operational issues gave rise to the need for a Maximum Power Point Tracker (MPPT) in order to maximize the power extraction from the needed renewable sources, on the one hand, and Flexible Power Point Trackers (FPPT) in order to meet the grid constraints due to the added PV sources, on another. Nevertheless, the MPP is a weather-dependent parameter and the tracking of which is governed by a convergence speed faster than the dynamic change of irradiance, hence the need for online techniques. The work in this document attempts to address this shortcoming by utilizing online gradient algorithms for this purpose. Numerical analysis and verification are presented hereunder, while the code of the algorithms can be found at \href{https://github.com/muhi-zatar/Online_gradient_descent_FPPT}{this link}.
\end{abstract}


\section{Introduction}
Due to the rapidly increasing economies and businesses, in addition to the need for grid decarbonization to avoid the catastrophic consequences of climate change, renewable energy resources have witnessed a dramatic expansion in the power-generating sector. Solar energy grew faster than other renewable energy sources, owing to its availability and minimized environmental effects. In addition, due to the reduced cost of photovoltaic (PV) panels, the growth rate of installation of PV power plants (PVs) is rapidly increasing around the globe \cite{shen2021facilitating, staffell2018increasing}.

The essential issue for both utility and residential PV owners is maximizing their revenue, which depends on cost and efficiency. Thus it is necessary to maximize the extracted power. However, the maximum power that can be extracted from the PV is a function of irradiance, temperature, and load; therefore, the maximum operating power differs with time and is unique. The variation in irradiance, temperature, or load changes the voltage at which maximum power transfer occurs, hence the need for an algorithm that continuously tracks the maximum power point and operates the system at this point. These algorithms are known as Maximum Power Point Trackers (MPPT) \cite{batarseh2018hybrid, khalane2017literature}, in which a DC-DC converter is implemented with a control scheme that continuously updates the duty cycle in order to match the PV source resistance to the load resistance, as seen by the source.

However, the increased penetration of renewable energy into the grid and power network poses new operational challenges that are primarily related to the stability and reliability of the power system. Therefore, new constraints and operational regulations are being implemented to fulfill the future's grid stability and power quality requirements. In order to address the operational challenges and limit the power injected from the PV into the grid, a new concept known as Flexible Power Point Tracking (FPPT) is presented. Figure 1 \cite{tafti2020extended} depicts how the power extracted from the PV is adjusted to a flexible power point that is regulated with grid constraints.
Several algorithms to track the FPPT are being developed, each with its advantages and disadvantages. One is a fast and reliable approach that takes a few iterations to find a local maximum for a given I-V curve, known as the Online Projected Gradient Descent (OPGD). This paper proposes using the OPGD technique for micro-inverter-based topology, as shown in Figure 2 \cite{zeb2018comprehensive}.

\begin{figure}[htbp]
\centerline{\includegraphics[width=8cm]{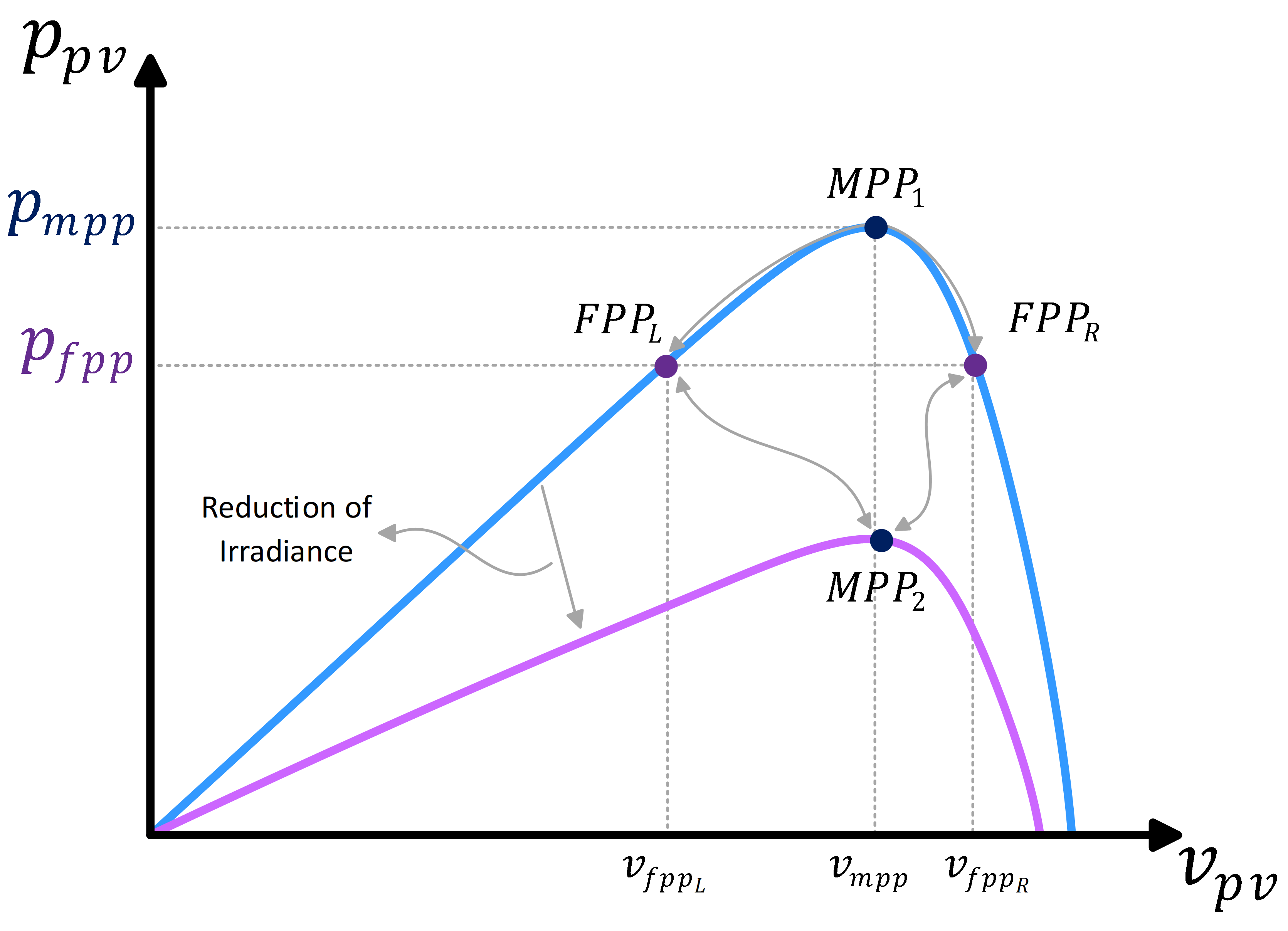}}
\caption{Maximum and Flexible power point tracking in PVs (MPP -Maximum power point FPP - Flexible power point) \cite{tafti2020extended}}
\label{fig}
\end{figure}

The rest of the document is divided as follows: section II lists related work and the need for an online algorithm to solve the problem of FPPT. Section III illustrates the problem formulation, whereas section IV presents how OPGD tracks the flexible power point. Section V lays out the simulation setup and the results. Finally, section VI presents the conclusion and practical implications of this work.

\begin{figure}[htbp]
\centerline{\includegraphics[width=8cm]{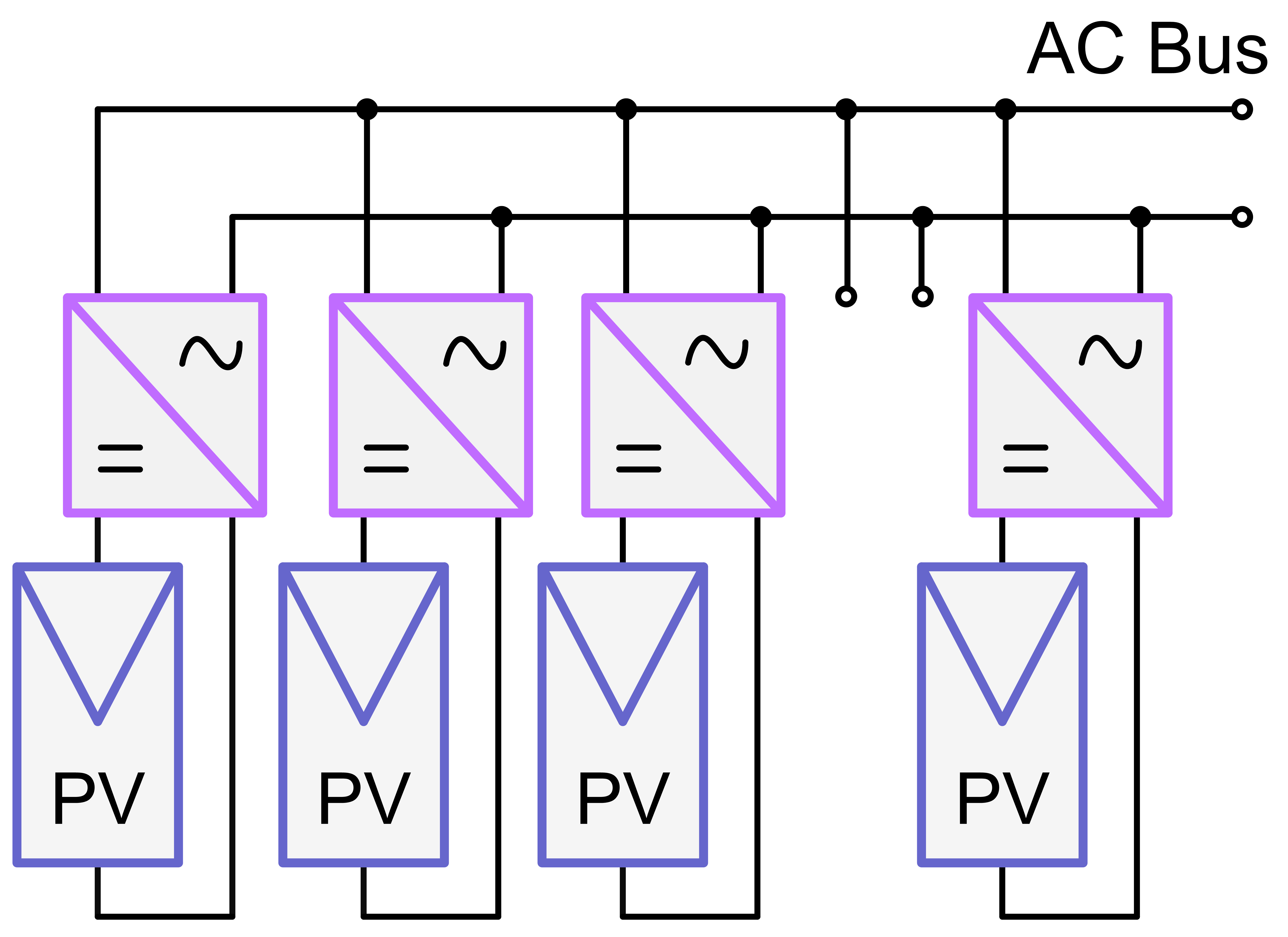}}
\caption{AC modules (micro-inverter) topology \cite{zeb2018comprehensive}}
\label{fig}
\end{figure}

\section{Time-Varying Constraints and Related Work}

\subsection{Time-Varying Constraints}
Depending on the level of penetration from inverter-based resources and the stability and reliability of the grid, countries have various regulations and standards for the operation of PV systems under different conditions. Each operational constraint imposes different constraints that vary with time, weather conditions, load, and other operational factors. The operational constraints are as follows \cite{tafti2020extended, peng2020grid}:

\begin{enumerate}
  \item \textbf{MPPT control zone (unconstrained)}: MPPT operation is implemented here, in which there are no operational constraints. The purpose of the algorithm in this zone is to extract the maximum available power from the PVs.

  \item \textbf{Power reserve constraint (delta power constraint)}: A delta power constraint is used to limit the active power from a PV to a desired constant value in proportion to the maximum available power. This operational zone is vital as a regulating reserve.

  \item \textbf{Ramp rate constraint}: A ramp rate constraint is used to limit the power ramp rate by which the active power can be changed

  \item \textbf{Power limiting control (absolute power constraint)}: An absolute power constraint is typically used to protect the power system against overload in critical situations.

\end{enumerate}

\subsection{Related Work and The Need For Online Optimization}

The maximum Power point tracking domain is older and hence more extensively explored than Flexible Power Point Tracking (FPPT). Therefore, to prove that FPP and MPP tracking need online optimization, MPPT literature will be considered as it is richer and better documented. There are numerous MPPT algorithms; each has its strengths and drawbacks. However, the majority of the literature only tests these algorithms under rapidly changing weather conditions (mainly irradiance) and only focuses on accuracy and convergence time. It is worth mentioning that irradiance variability is around $80 \mathrm{~W} / \mathrm{m}^{2}$ in less than 2 seconds \cite{lave2010solar}, \cite{jain2020quantifying} and varies more with larger time steps. An $80 \mathrm{~W} / \mathrm{m}^{2}$ change can contribute to a $6 \%$ loss in the power extracted if the operational point remains unchanged. In addition to the weather, the load is continuously changing, affecting the optimal operation point. The algorithms in literature usually consume more time to converge to the maximum power point, which is summarized in Table 1\cite{bradai2017experimental}:

\begin{table}[htbp]
\caption{T HE PERFORMANCE OF ALGORITHMS IN TERMS ON CONVERGENCE SPEED}
\begin{tabular}{ccc}
\hline
\textbf{Algorithm}                      & \textbf{\begin{tabular}[c]{@{}c@{}}Best \\ Convergence\end{tabular}} & \textbf{\begin{tabular}[c]{@{}c@{}}Worst\\ Convergence\end{tabular}} \\ \hline
Fuzzy Logic Controller (FLC) \cite{alajmi2011maximum}  & 2.4 sec                                                              & 5.4 sec                                                              \\
Particle Swarm Optimization (PSO) \cite{lian2014maximum} & 6.59 sec                                                             & 9.69 sec                                                             \\
Switched Photovoltaic {}\cite{elserougi2015switched}             & 11.6 sec                                                             & 16.8 sec                                                             \\
Artificial Bee Colony (ABC) \cite{sundareswaran2014enhanced}       & 5.01 sec                                                             & 5.63 sec                                                             \\ \hline
\end{tabular}
\end{table}
Therefore, as can be noted from Table 1, since the convergence time is less than the time it takes for the environmental conditions to change, there is a pressing need for an online algorithm.

\section{Problem Formulation}
As aforementioned, FPPT aims to regulate the overall maximum power extracted from the PV modules under different grid, load, and operational constraints. However, the maximum power is continuously affected by the irradiance and temperature. In contrast, the flexible power is governed by the operational constraints and load, which affect the regulated maximum operation point of the PV. In this work, partial shading conditions are not studied as they have no significant impact under the micro-inverter topology adopted in Figure 2.

The power extracted from a PV is given by equation (1):

\begin{equation}
    P_{p v}=V * I_{L}-V * I_{o} * e^{\frac{V}{V_{t}}}  
\end{equation}

Where $V$ is the PV voltage, $I_{L}$ is the light-generated current, $I_{o}$ is the saturation current, and $V_{t}$ is the thermal voltage. The light-generated current is a function of irradiance. In contrast, the thermal voltage $V_{t}$ varies with the module temperature, which changes the power extracted and the required voltage $V$ to operate at the maximum power point.

For this work, in order to treat the problem as a minimization instead of a maximization problem, the power will be multiplied by $-1$. The flow of analysis is inspired by the work in \cite{selvaratnam2018numerical}

\subsection{Preliminaries}
Let $N:=\{1,2, \ldots\}$ and $N_{0}:=N \cup\{0\}$. In this problem, $F$ is considered the overall power in which $F:=\left\{f_{t}: R^{n} \mid k \in\right.$ $\left.N_{0}\right\}$ represents a family of continuous functions and where $X:=\left\{X_{t} \subset R^{n} \mid t \in N_{0}\right\}$ are closed, non-empty sets that capture the possible time-varying constraints, and where $n$ represents the number of PV modules that are connected to the grid. The algorithm is required to generate solutions to the following sequence of optimization problems as in equation $(2)$, using iteration of the form shown in equation (3):

\begin{equation}
\min _{x \in X_{t}} f_{t}(x)
\end{equation}

\begin{equation}
x_{t+1}=G_{t}\left(x_{t}\right)
\end{equation}

Where the operator $G_{t}: R^{n} \rightarrow R$ uses first-order information and knowledge of the feasible set $X_{t}$ at the current time step.

In this work, uniform irradiance across the PV panels is assumed. Hence, it is safe to say that attention is restricted to families of locally smooth, strongly convex functions with Lipschitz gradient and convex constraints, and hence the following assumptions:

\textbf{Theorem 1 (Smoothness)}: Every $f \in F$ is twice continuously differentiable. This indicates that the power is an exponential function that is known to be smooth and is not globally Lipschitz continuous

\textbf{Theorem 2} (Uniform strong convexity): It is well documented and known in the literature that a PV under uniform irradiance has one unique maximizer (minimzer) and, therefore, is strongly convex.

\textbf{Assumption 1} (Convex Constraints): Every $X \in X$ is closed, convex and non-empty.

This work employs the online projected gradient descent algorithm to solve the problem presented in equation (2), which applies a specific choice for $G_{t}$ that is shown in equation (3). Under assumption 1, $f_{t}$ is guaranteed to have a unique global minimizer over $X_{t}$.

\subsection{Metrics}
In this work, The following metrics will be used to evaluate the performance of the algorithm:

\begin{enumerate}
  \item \textbf{Tracking Error}: $\left\|x_{t}-x_{t}^{*}\right\|$, which represents the difference between the estimated voltage and the actual voltage at MPP.

  \item \textbf{Instantaneous regret}$f_{t}\left(x_{t}\right)-f_{t}\left(x_{t}^{*}\right)$ which represents the difference between the output power and the maximum available power at time $t$ and hence it represents the power lost. In the context of FPPT, the instantaneous power is greater than zero.

  \item \textbf{Average Dynamic Regret}$\sum_{t=1}^{T} f_{t}\left(x_{t}\right)-\sum_{t=1}^{T} f_{t}\left(x_{t}^{*}\right)$ which represents the average power between the algorithms resulting output power and the maximum power available. It is useful to compute the dynamic regret as if multiplied by the operational time, representing the overall energy loss; therefore, it is important to compute the dynamic regret.

  \item \textbf{Static Regret}: $\sum_{t=1}^{T} f_{t}\left(x_{t}\right)-\min _{x \in X} \sum_{t=1}^{T} f_{t}\left(x_{t}^{*}\right)$ Since there exists a tracking algorithm that is called constant voltage which fixes the operational voltage at a specific value and does not change it. Therefore it is important to compute the static regret.

\end{enumerate}

The instantaneous regret is defined as in equation (4):

\begin{equation}
    \phi_{t}=f_{t}\left(x_{t}\right)-f_{t}\left(x_{t}^{*}\right)
\end{equation}

\begin{center}

\end{center}

\subsection{Unconstrained Case}
First of all, bounds for the unconstrained case (MPPT control) are considered. The guarantee that $\nabla f_{t}\left(x_{t}^{*}\right)=0$ when $X_{t}=R^{n}$ leads to the inequality in (5):

\begin{equation}
f_{t}\left(x_{t}\right)-f_{t}\left(x_{t}^{*}\right) \leq \frac{L}{2} e_{t}^{2} 
\end{equation}

Consequently, instantaneous regret is bounded as following when $\alpha=\frac{2}{L+\mu}$ which is the value used to minimze the upper bound, then:

\begin{equation}
lim _{t \rightarrow \infty} \phi_{t} \leq \frac{L V^{2}(L+\mu)^{2}}{8 \mu^{2}}
\end{equation}

\subsection{Constrained Case}

The previous error bounds apply equally well to constrained and unconstrained problems. However, the presence of constraints means that extra consideration should be considered. The bounds in the previous subsection do not apply as the minima does not have to be a stationary point.

Finite time feasibility: the following assumption of finite time feasibility guarantees the feasibility of the projected gradient descent as follows:
\begin{equation}
   \left\|x_{t+1}-x_{t}^{*}\right\| \leq D 
\end{equation}

Where $D<+\infty$

Bounds on the temporal variability:

\begin{equation}
    \left\|x_{t+1}^{*}-x_{t}^{*}\right\| \leq \frac{V}{\mu}
\end{equation}

Proof: To show the computation of the bound, starting from (derived in the previous problem):

\begin{equation}
   \left\|x_{t}-x_{t}^{*}\right\| \leq\left\|x_{t}-x_{t-1}^{*}\right\|+\left\|x_{t-1}^{*}-x_{t}^{*}\right\| 
\end{equation}

Where $f_{t}$ is strongly convex with $\mu$.

Now substituting $x=x_{t+1}^{*}$ and using the fact that $\nabla f_{t+1}\left(x_{t+1} *\right)=0$ and adding it to the above, the equation becomes:

$$
\left\|\nabla f_{t+1}\left(x_{t+1}^{*}\right)-\nabla f_{t}\left(x_{t+1}^{*}\right)\right\| \geq \mu\left\|x_{t+1}^{*}-x_{t}^{*}\right\|
\ \ \ \ \ (10)$$

And assuming that:

\begin{equation}
    \left\|\nabla f_{t+1}\left(x_{t+1}^{*}\right)-\nabla f_{t}\left(x_{t+1}^{*}\right)\right\| \leq V
\end{equation}

Then the bound is:
\begin{equation}
    \left\|x_{t+1}^{*}-x_{t}^{*}\right\| \leq \frac{1}{\mu}\left\|\nabla f_{t+1}\left(x_{t+1}^{*}\right)-\nabla f_{t}\left(x_{t+1}^{*}\right)\right\| \leq \frac{V}{\mu}
\end{equation}

Uniform Lipschitz cost for such $G \geq 0$, there exists: For the pair $F, X$ there exists $G>0$ where

\begin{equation}
 \|\nabla f_{t}(x) \|\ \leq G   
\end{equation}

\section{ALGORITHMS}
This section presents the algorithms that attempted to solve the FPPT problem explained earlier.

\subsection{Proposed Online Algorithm}\label{AA}

This work proposes using the online optimization method, Online Projected Gradient Descent (OPGD). First, the sensors in the system take different relative measurements at each time; irradiance, temperature, and load-related values. As in (1), the power equation can be calculated with sufficient measurements. Thus, the gradient for the voltage can be calculated to get the optimal operating voltage. Finally, the DC-DC converter's duty cycle is calculated according to the optimal value that equalizes the load resistance as seen by the source resistance to the source resistance and thus achieving maximum power transfer. Figure 3 illustrates the flow of the algorithm.
Online projected gradient descent will be executed with a constant step size of $\alpha = \frac{2}{10}$.

\begin{figure}[htbp]
\centerline{\includegraphics[width=7cm, height=5cm]{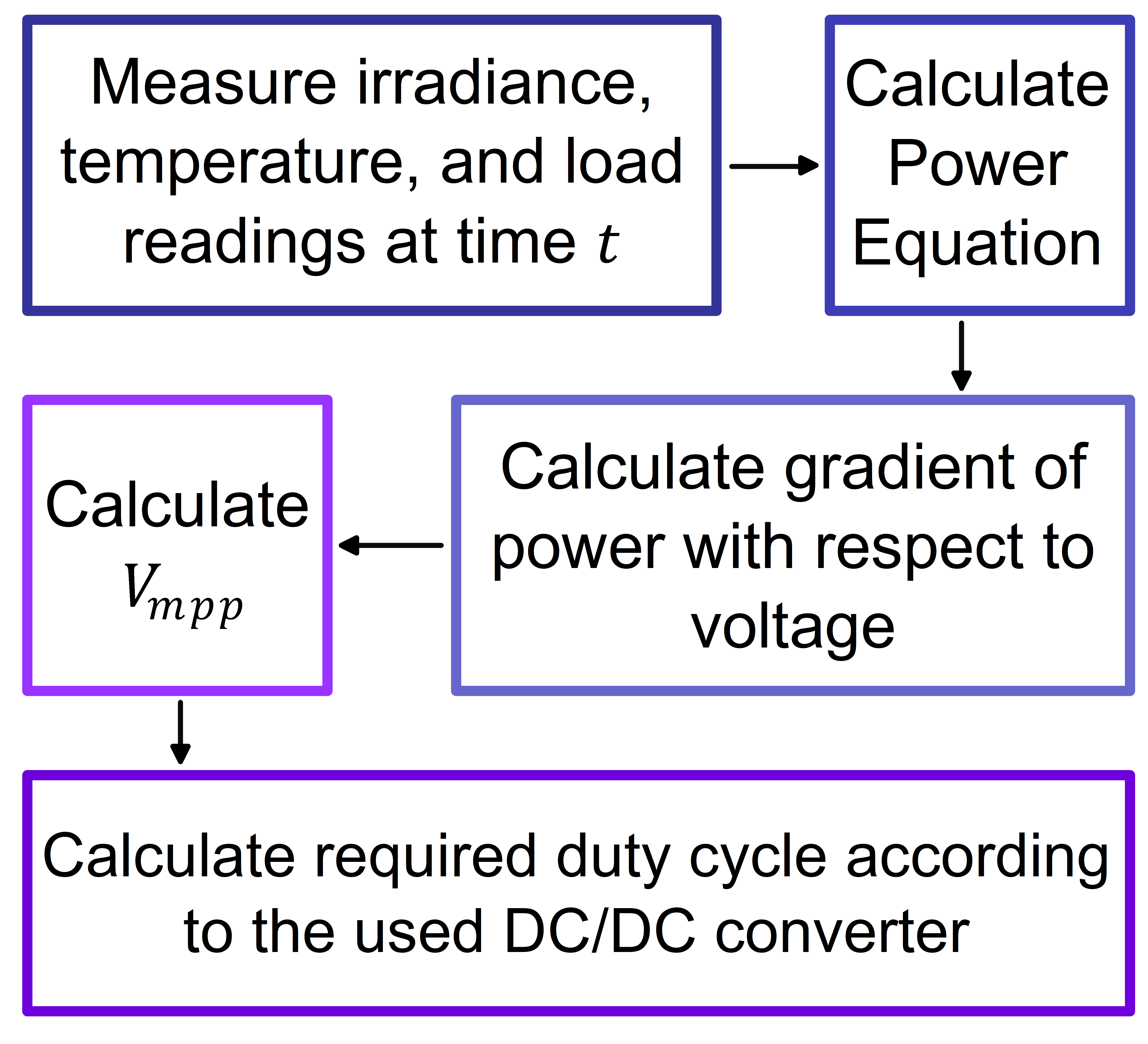}}
\caption{Online projected gradient descent algorithm flow}
\label{fig}
\end{figure}

\subsection{Perturb and Observe}

Perturb and Observe (P\&O) is the most popular and widely used tracking algorithm for PV systems due to its relatively good accuracy, ease of implementation, and being PV module independent. This work compares the performance of online gradient descent to the widely used P\&O as a benchmark.

\section{Simulation Results and Discussion}
This section presents the simulation setup and associated simulation results.

\subsection{Simulation Setup}

MATLAB Simulink is used as a simulation tool to build the following system configuration shown in figure 4.

\begin{figure}[htbp]
\centerline{\includegraphics[width=8cm, height=3cm]{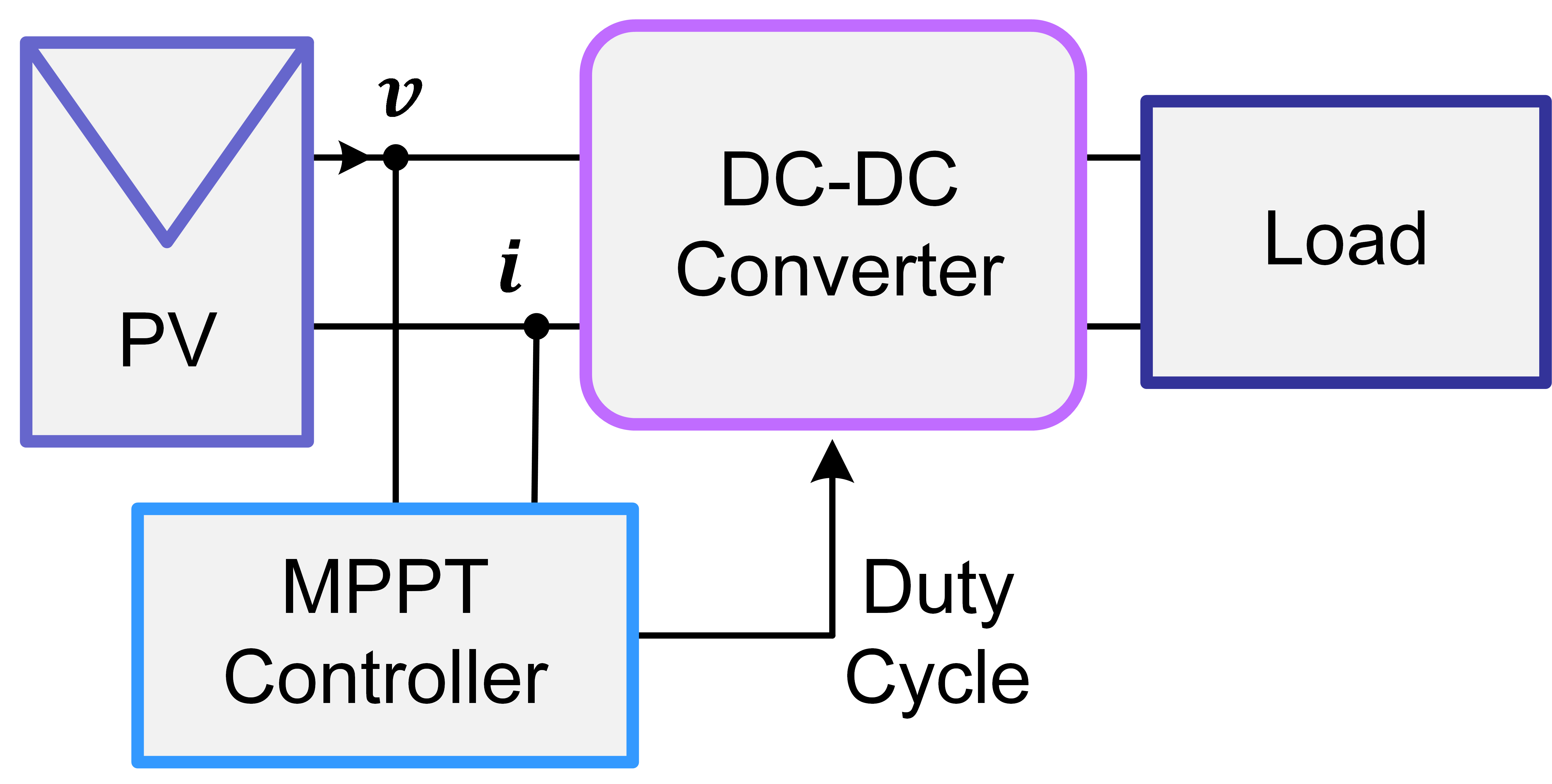}}
\caption{Maximum (flexible) Power Point block diagram}
\label{fig}
\end{figure}

It is worth mentioning that for flexible power point tracking, the same controller will be used but with constraints on the
operating conditions and thus has the same block diagram as above.

Table 2 summarizes the characteristics of the PV modules
used.

\begin{table}[h]
\centering
\caption{PV MODULES CHARACTERISTICS}
\begin{tabular}{ll}
\hline
\textbf{PV Characteristic} & \textbf{Value} \\ \hline
Number of modules          & 8              \\
Open Circuit Voltage       & 36.3 V         \\
Short Circuit Current      & 7.84 A         \\
MPP Voltage at STC         & 29.2 V         \\
MPP current at STC         & 7.3 A          \\
Maximum Power at STC       & 213.15 W       \\ \hline
\end{tabular}
\end{table}

Finally, the algorithms were tested under a highly fluctuating irradiance to test their ability to track the optimal operating point. The code and simulation files are available at this link.

\subsection{Simulation Results}

\subsubsection{Unconstrained Case:}

Figures 5 and 6 below show the plots of the tracking and instantaneous errors respectively for the online gradient descent algorithm (the P\&O graphs for clarity). Each time t is a 0.1 second step.

\begin{figure}[htbp]

\centerline{\includegraphics[width=8cm]{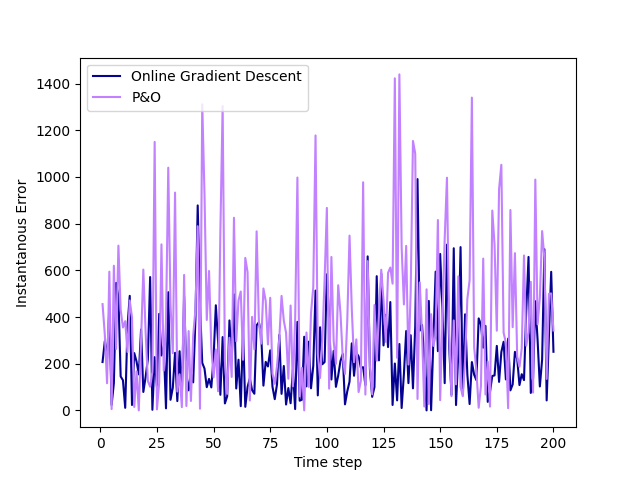}}
\caption{Instantaneous error - Unconstrained case}
\label{fig}
\end{figure}

\begin{figure}[htbp]
\centerline{\includegraphics[width=8cm]{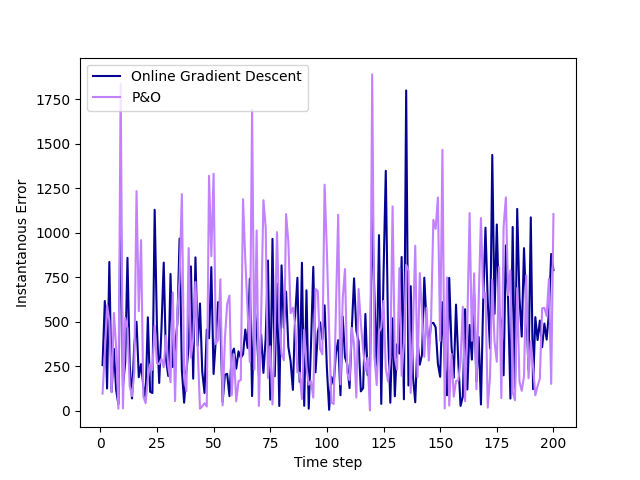}}
\caption{Tracking error - Unconstrained case}
\label{fig}
\end{figure}

For further evaluation and comparison, the average dynamic regret and static regret are calculated for both the online gradient descent and P\&O. For static regret, the operating voltage was simply fixed at 29 V. Table 3 presents the results for each algorithm. The dynamic regret shown is at the end of the simulation (at time T).

\begin{table}[h]
\caption{OGD AND P\&O COMPARISON FOR THE UNCONSTRAINED CASE}
\centering
\begin{tabular}{lll}
\multicolumn{1}{c}{\textbf{Metric}} & \multicolumn{1}{c}{\textbf{OGD}} & \multicolumn{1}{c}{\textbf{P\&O}} \\ \hline
Average Dynamic Regret              & 54.67                            & 67.3                              \\
Static Regret                       & 51.36                            & 55.3   \\ \hline            
\end{tabular}
\end{table}

\subsubsection{Constrained Case}

As aforementioned, sometimes the grid imposes some operational challenges that require some actions from the PV controllers, such as solar curtailment, power control, or over-voltage protection, which in turn imposes constraints on the PV controllers. In the two constrained cases, the operational mode is varied between the MPPT case (unconstrained), over-voltage protection, and active power control. Figures 7 and 8 below show the plots of the tracking and instantaneous errors, respectively, for the online gradient descent algorithm in the unconstrained case. Each time t is a 0.1-second step.

\begin{figure}[htbp]
\centerline{\includegraphics[width=8cm]{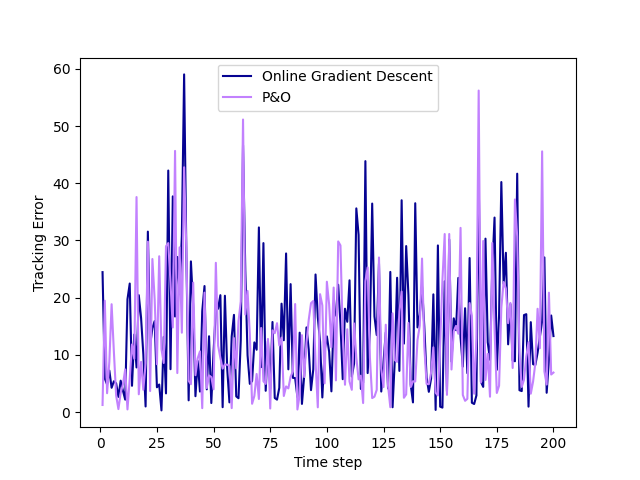}}
\caption{Instantaneous error - Constrained case}
\label{fig}
\end{figure}

\begin{figure}[htbp]
\centerline{\includegraphics[width=8cm]{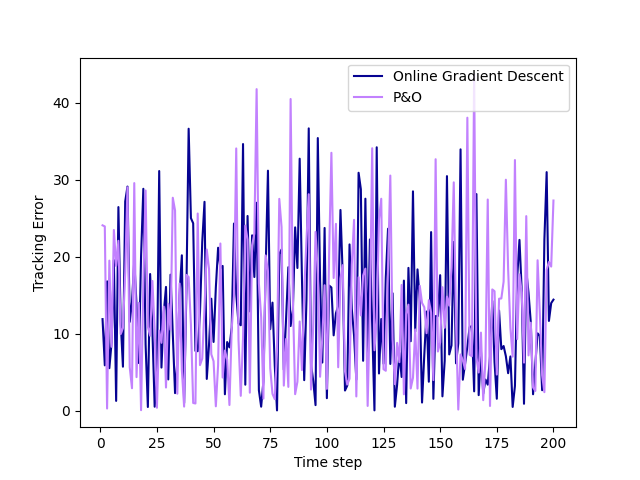}}
\caption{Tracking error - Constrained case}
\label{fig}
\end{figure}

Similar to the unconstrained case, the average dynamic and static regrets are calculated and compared against the P\&O as shown in table 4. In the case of static regret, if the fixed operating voltage is above
the constraint, it is reduced to the allowed value. The dynamic regret shown is at the end of the simulation (at time T).

\begin{table}[htbp]
\caption{OGD AND P\&O COMPARISON FOR THE CONSTRAINED CASE}
\centering
\begin{tabular}{ccc}
\textbf{Metric}        & \textbf{OGD} & \textbf{P\&O} \\ \hline
Average Dynamic Regret & 61.363       & 72.4          \\
Static Regret          & 55.1         & 57.41         \\ \hline
\end{tabular}
\end{table}

\section{Conclusion and Practical Implications}

As there are 8 PV modules, with each module able to generate 213.15 W under Standard Test Conditions (STC)
(1000 W/m 2 irradiance and 25 C temperature). The average loss of power for OGD for the unconstrained case is 54.7
W and 61.36 for the constrained case. The P\&O algorithm performance was relatively worse, achieving 67.3 W
and 72.4 W loss of power in the unconstrained and constrained cases, respectively. However, in the span of time, the energy
yielded from the OGD will be considerably higher than the energy harvested by the P\&O. Also, it has been noticed that
when the sudden change in irradiance or load is high, the P\&O performance becomes significantly worse, as it has to search
for the MPP, which is usually much further from the original operating point. On the other hand, OGD is not affected by the
variability in irradiance, as it calculates the required duty cycle and operating voltage rather than searching for it,
Therefore, the OGD is better at placing the FPP under extremely varying conditions.

\bibliographystyle{IEEEtran}
\bibliography{references}

\end{document}